\documentclass[aps,twocolumn,amsmath,amssymb,a4paper,9pt,showkeys,notitlepage]{revtex4}

\def\pdftitle{ParticleEncapsulation}
\def\authorname{Leon Abelmann}
\def\pdfsubject{}
\def\pdfkeywords{}
\def\pdfbackref{none}

\usepackage[pdftex,
        colorlinks=true,
        urlcolor=darkblue,               
        anchorcolor=darkblue,
        filecolor=green,                   
        linkcolor=darkblue,              
        menucolor=darkblue,
        citecolor=darkblue,
        pagebackref,
        backref={\pdfbackref},
        pdfpagemode={UseOutlines},
        bookmarks=true,
        bookmarksopen=true,
        pdftitle={\pdftitle},
        pdfauthor={\authorname}, 
        pdfsubject={\pdfsubject},
        pdfkeywords={\pdfkeywords}
        ]{hyperref}

\usepackage{graphicx}
\DeclareGraphicsExtensions{.pdf,.png,.jpg}
\usepackage{thumbpdf}
\usepackage{color}
\definecolor{darkgreen}{rgb}{0,0.5,0}
\definecolor{darkblue}{rgb}{0,0,0.5}
\definecolor{brown}{rgb}{0.98,0.92,0.73}
\definecolor{red}{rgb}{1,0,0}
\definecolor{yellow}{rgb}{1,1,0}
\definecolor{blue}{rgb}{0,0,1}
\definecolor{green}{rgb}{0,1,0}
\definecolor{purple}{rgb}{1,0,1}
\definecolor{gray}{rgb}{0.8,0.8,0.8}
\definecolor{black}{rgb}{0,0,0}
\definecolor{white}{rgb}{1,1,1}
\definecolor{gold}{rgb}{1.,0.84,0.}

\graphicspath{{~/Documents/Images//}{Figures//}}

\usepackage{amsmath}
\usepackage{wasysym}

\usepackage{booktabs}

\usepackage{lineno}

\usepackage{pdfsync}

\usepackage{doi}

\usepackage{siunitx}

\usepackage{nicefrac}

\def\bs{\boldsymbol}

\def\bs{\boldsymbol}

\newcommand{\grad}[1]{%
        \bs{\nabla} #1
}

\usepackage[paper=a4paper,total={170mm,241mm},top=20mm,left=20mm,columnsep=8mm]{geometry}\usepackage{siunitx}
\usepackage{nicefrac}

\def\figurewidth{0.8\columnwidth}
\def\widefigurewidth{\columnwidth}

\newif\ifcmtr
\cmtrtrue
\ifcmtr 
\newcommand{\cmtr}[1]{ %
   [\color{red} \textbf{#1} \normalcolor]%
}%
\else
\newcommand{\cmtr}[1]{ %
}%
\fi

\newif\ifcmtrj
\cmtrjtrue%
\ifcmtrj 
\newcommand{\cmtrj}[1]{ %
   [\color{green} \textbf{#1} \normalcolor]%
}%
\else
\newcommand{\cmtrj}[1]{ %
}%
\fi

\begin{document}
\title{Absence of enhanced uptake of fluorescent magnetic particles
  \\into human liver cells in a strong magnetic field gradient}

\date{\today}

\author{Leon Abelmann$^{1,2,3}$}
\author{Eunheui Gwag$^1$}
\author{Baeckkyoung Sung$^1$}
\affiliation{}
\affiliation{ 
  $^1$KIST Europe, Saarbr\"ucken, Germany\\
  $^2$University of  Twente, The Netherlands\\
  $^3$Saarland University, Saarbr\"ucken, Germany
\underline{l.abelmann@kist-europe.de}
}

\begin{abstract}
  We investigated whether we can detect enhanced magnetic nanoparticle
  uptake under application of a large magnetic force by tagging the
  particles with a fluorescent dye. Human liver cells were cultured in
  a micro-channel slide and exposed to two types of magnetic
  nanoparticles with a diameter of \SI{100}{nm} at a concentration of
  \SI{10000}{particles/cell} for \SI{24}{hours}. Even though we
  achieved a magnetic force that exceeded the gravitational force by a
  factor of \num{25}, we did not observe a statistically significant
  increase of immobilised particles per cell.
\end{abstract}

\maketitle 

\section{Introduction}
The interaction of micro- and nanoparticles with cells is a
challenging research area of major importance. Micro- and
nanoparticles, which may be
toxic~\cite{Elsaesser2012}, can enter our body by accident,  or they can be administered intentionally in
biomedical procedures such as drug
delivery~\cite{Anselmo2016,Mornet2004} and
\textit{in vivo} imaging~\cite{Pankhurst2003, Zabow2011}.

Interaction studies between nanoparticles and cells are mainly
performed using \textit{in vitro} cell cultures in multi-well
plates~\cite{Soenen2010}. In the case of small molecules, diffusion
ensures that the molecule concentration is reasonably constant over
the volume of the well. However, micro- and nanoparticles are subject
to sedimentation. This has two implications. First, sedimentation gradually increases particle concentration at the cell membrane, the rate of which depends strongly on the particle
diameter. Secondly, the particles exert a force on the cell membrane,
which may affect particle incorporation~\cite{Teeguarden2006}.

Magnetic micro- and nanoparticles have the advantage that they can be
manipulated by external magnetic fields, which is exploited in
targeted drug delivery~\cite{Pondman2014, Kim2010},
mechano-stimulation~\cite{Kilinc2016} and hyperthermia
treatment~\cite{Colombo2012}. The magnetic forces that one can apply
are orders of magnitude greater than gravitational forces. Therefore, by magnetically attracting nano-particles towards the
bottom of the well, we can accelerate sedimentation and increase
particle incorporation.

There have been a few reports on the increase of particle
uptake. For example, Prijic and colleagues elegantly demonstrated the increased
uptake of super-paramagnetic particles~\cite{Prijic2010}. They found that the total
iron content in the cell, measured by inductively coupled plasma
atomic emission spectroscopy, increased by a factor of
\num{3}--\num{8}. Unfortunately, this method  requires one million or more cells. Moreover, one cannot
discriminate between the uptake of particles or iron ions in
the solution. The same researchers also observed particle uptake by
transmission electron microscopy. This method requires fewer cells, but
is very labour-intensive. It is also difficult to discern between particles
on the cell membrane and those incorporated in the cell.

Rather than observing the particles themselves, one can observe their
effect on the cell. One elegant option is to use magnetic particles to
transinfect cells, a process called magnetofection~\cite{Haim2004, Scherer2002}. Pickard and Chari~\cite{Pickard2010} demonstrated this by attaching green
fluorescent protein (GFP) plasmids to Neuromag SPIONs. When the
particles entered the cell, the plasmids were reproduced. The
subsequent generation of GFP determined whether cells are
transfected. The application of force by means of magnetic field
gradients enhanced the uptake by a factor of 5. A slow oscillation
seemed to have a positive effect.

Particles in a cell can be identified with optical microscopy by means of fluorescence, for which Dejardin and colleagues used ScreenMag-Amine
magnetic particles tagged with
fluorescein~\cite{Dejardin2011}. The particles were treated with
activated penetratin to increase their uptake. By integrating the
intensity of the emitted light over the sample area, an increase in
uptake of about \SI{30}{\percent} was observed. The magnet used in
this experiment was only \SI{13}{mm} in diameter, leading to particle
accumulation in the center of the observation area and subsequent loss
of information on particle concentration. A similar approach was taken
by Venugopal and colleagues~\cite{Venugopal2016} as well as by Park and
colleagues~\cite{Park2016}, who additionally showed by flow cytometry
an increase in uptake ranging from a factor of 0.5 (Venugopal) to 7 (Park).

\begin{figure}
    \centering
    \includegraphics[width=\widefigurewidth]{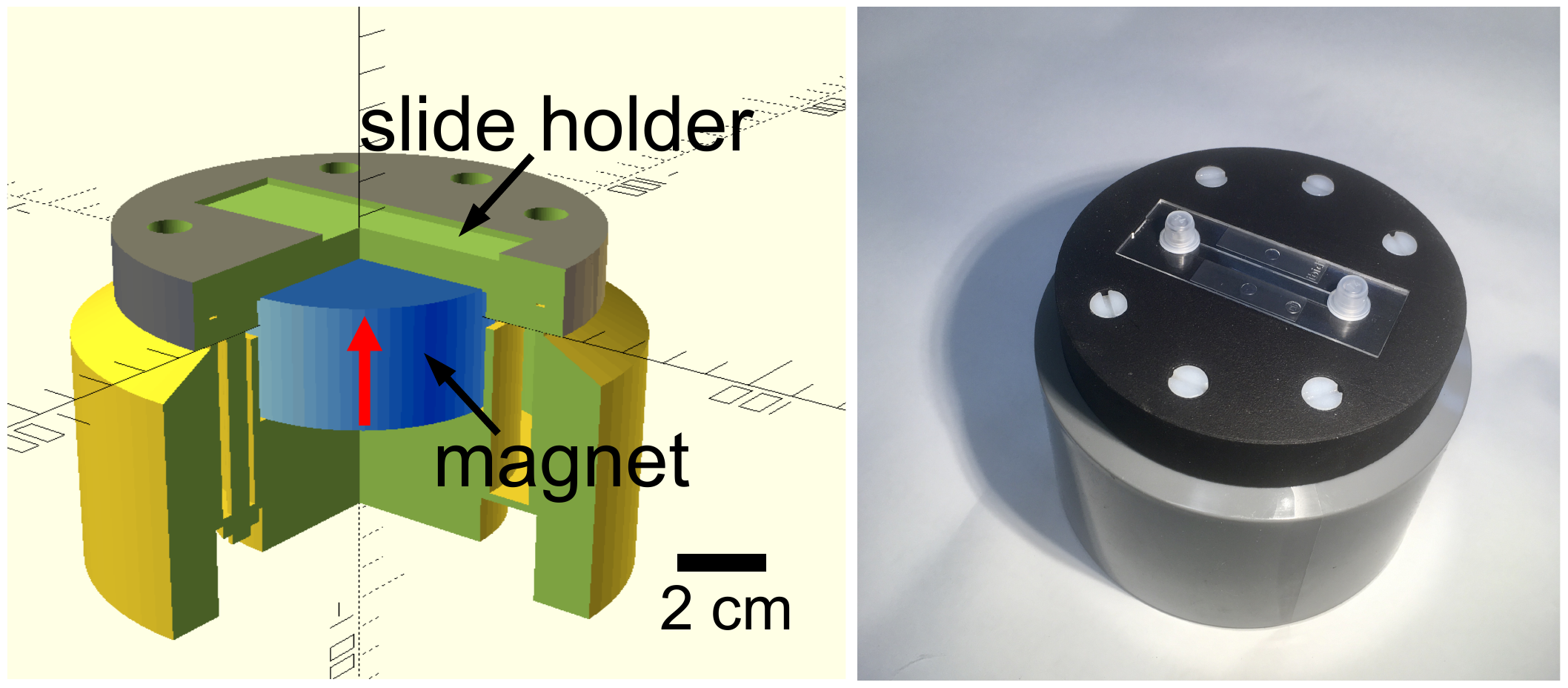}
    \caption{Holder to investigate the effect of a high magnetic field
    gradient on the incorporation of magnetic nanoparticles into cells.}
    \label{fig:IbidiHolder}
\end{figure}

Encouraged by these fluorescence experiments, we
investigated whether one can use optical microscopy to observe
individual fluorescent magnetic nanoparticles in \textit{in vitro} cell
cultures to detect an increase in magnetic particle uptake under
application of a magnetic force.
For this experiment, we designed a system with a large magnet of
\SI{70}{mm} diameter, on top of which Ibidi $\mu$-Slide channel slides
could be mounted (Figure~\ref{fig:IbidiHolder}). Using such a big
magnet ensured that the forces are strong and uniform. Human liver cells from a HepG2
cell line in the channel
slides were exposed to fluorescent magnetic nanoparticles of
\SI{100}{nm} diameter. The number of particles per area served as a
metric to study the influence of the magnetic force. We conclude that our experimental configuration showed no
significant effect of the magnetic field.

\section{Theory}
The force on a magnetic object that is small compared to the spatial
variation of the externally applied field $\bs{B}$ [T] can be
approximated from its total magnetic moment $\bs{m}$ [A/m]

\begin{eqnarray}
  \label{eq:grad}
  \bs{F} = -\grad{(\bs{m}\bs{B})}
  \text{.}
\end{eqnarray}
%
The magnetic moment of a magnetic object in a fluid is generally a
function of strength, direction and history of the applied field. The
applied field is a combination of the external field and the field of
all other magnetic particles in the fluid. Moreover, very small
particles will be subject to Brownian motion. Therefore, in principle, the
calculation of forces on magnetic particles in a magnetic field
gradient is complex. To obtain first approximations, we consider a
single particle that is either a permanent magnetic dipole or a soft
magnetic sphere.

In the case of the permanent magnetic dipole approximation, we assume a particle
with a permanent magnet moment $\mu_0m_\text{r}=I_\text{r}V_\text{p}$,
where $I_\text{r}$ [T] is the remanent magnetisation of the particle
with volume $V_\text{p}$ [m$^3$]. We further assume that field changes are
slow such that particles can rotate against viscous drag into the
direction of the field. In this case, equation~(\ref{eq:grad}) reduces
to

\begin{eqnarray}
  \label{eq:magnet}
  \bs{F}=m_\text{r} \grad{B}\text{, } B=|\bs{B}| \text{.}
\end{eqnarray}

For the second approximation, we assume a soft
magnetic sphere with a susceptibility of $\chi$, which is the ratio between
the magnetisation $I$ [T] in the particle and the internal field
$B_\text{in}$ [T]. As a sphere has a demagnetisation factor of
1/3,  the internal magnetic field is 

\begin{equation*}
  \label{eq:chi}
  B_\text{in}=-B + \frac{1}{3}I = -B + \frac{1}{3}\chi
  B_\text{in} 
\end{equation*}
where
\begin{equation*}  
  I = \frac{3\chi}{(3+\chi)}B 
\end{equation*}
and all fields are (anti-)parallel. In this approximation, the energy
and resulting force are

\begin{equation}
\begin{aligned}
  \label{eq:3}
  U &= -\frac{1}{2}V_\text{p}\frac{3\chi}{\mu_0(3+\chi)}B^2  \\
  F &= \frac{1}{2}V_\text{p}\frac{3\chi}{\mu_0(3+\chi)} \grad{B^2}  \\
  &=   V_\text{p}\frac{3\chi}{\mu_0(3+\chi)} (\bs{B}\grad{})\bs{B} 
  \text{.}
\end{aligned}
\end{equation}
The factor 1/2 originates from integrating from $-\infty$, where the
energy is 0, and we used the vector identity
$ \grad{B^2}=2 (\bs{B}\grad{})\bs{B}$.


\section{Experimental}
\label{sec:exp}
To maximise uniformity, we based the magnetic field system on the
largest NdFeB magnet we could readily obtain
(Supermagnete.de). This magnet has a diameter of \SI{70}{mm}, a height
of \SI{35}{mm}, and is made of N45, which is specified to have a
remanent magnetisation of \SI{1.3}{T}. These magnets can seriously injure the
experimentalist if accidentally brought too close
together. Therefore we encased them in PVC 
(\SI{70}{mm} (121605), \SI{100}{mm} (121457) and \SI{125}{mm} (121620)
reducer rings from Wildkamp, Netherlands). The position of and
distance between the $\mu$-Slide channel and the magnet were accurately
fixed using a 3D-printed nylon plastic top holder (see
Figure~\ref{fig:IbidiHolder}). The source files for the 3D-printed
holder are available as additional material.

To calculate the magnetic field and forces generated by the magnet, we
integrated the magnetic charge densities. In
contrast to finite-element methods, the field is  calculated only at
the points of interest, which is much faster at high
precision. The resulting equations are generated automatically by the
MagMMEMS package, which is a preprocessor for
Cades~\cite{Delinchant2007}.  The input files are available in the
supplementary material.

The magnetic field above the magnet is measured with a MetroLab THM1176
three-axis Hall sensor attached to a microscope glass slide
to obtain the field components at the same height as the $\mu$-Slide channel. To
visualise the particle density, we used Ibidi $\mu$-Slide I Luer channels
(Ibidi 80176) filled with bare iron-oxide nanoparticles of
\SI{5}{nm} diameter (EMG304 by FerroTec).

For cell studies, human hepatoma HepG2 cells (ATCC, HB-8065) were
cultured in Eagle's minimum essential medium supplemented with
\SI{10}{\percent} fetal bovine serum and \SI{1}{\percent}
penicillin-streptomycin in an incubator at \SI{37}{\celsius} and
\SI{5}{\percent} CO$_2$ atmosphere. The cell concentration was
determined by means of a hemocytometer and diluted to
\SI{5e4}{cells/mL}. Of this solution, \SI{30}{\micro L} was introduced
into Ibidi $\mu$-Slide VI 0.4 channels with an Ibitreat surface
coating to promote cell adhesion (Ibidi 80606). The dimension of the
channels is \SI{0.4x3.8x17}{mm}, hence the maximum cell density is
\SI{23}{cells/mm^2}.

The Ibidi channels were left inside the incubator for \SI{24}{hours}
on top of the holder with and without a magnet before analysis. To assess
cell viability, we used a live/dead double-staining assay (Sigma-Aldrich,
04511). Analysis was performed on \num{10} images chosen randomly
over the channel area. Average and standard deviations were
calculated from three independent experiments (\num{30}
images in total).

We studied the interaction between cells and magnetic nanoparticles with an
average diameter of \SI{100}{nm}. For this we used red fluorescent
cross-linked dextran iron-oxide cluster-type particles (94-00-102 from
Micromod) with a specified iron concentration of \SI{6}{mg/mL} and a
particle concentration of \SI{6e12}{particles/mL}. In addition, we used
green fluorescent (\SI{510}{nm}) iron oxide incorporated conjugated
polymer particles (905038 from Sigma-Aldrich) with a specified iron
concentration of \SI{100(10)}{\micro g/mL}. The composition of the
polymer and the particle concentration are not specified by the manufacturer.

Both suspensions were first diluted by a factor of 235 and added to the cell culture medium at a
volume ratio of 1:51. The final particle
concentration for the MicroMod particles in the cell culture is
therefore \SI{5e8}{particles/mL} with an iron concentration of
\SI{0.5}{\micro g/mL}. Hence, there is an average of \SI{1e4} particles per
cell in the medium. The final iron concentration for the cell culture
with Sigma-Aldrich particles is \SI{8}{ng/mL}.

Images were taken with a Leica DMi8 fluorescence microscope at a size
of \SI{2048x2048}{pixel}. Both a \num{20}$\times$ and a \num{40}$\times$ lens were
used, calibrated at \SI{324}{} and \SI{162}{nm/pixel}, respectively.

Particles were counted using ImageJ software (Wayne Rasband,
NIH, USA). The image taken with a red (Micromod particles) or green
(Sigma-Aldrich particles) filter was converted into 8-bit greyscale. An
intensity threshold was applied to create a 1-bit mask. The ``Analyze
Particles~...'' script was run to count the number of particles using a
lower size threshold. Both the intensity threshold and size threshold
were varied. The ImageJ script that automates this process and
generates the overlay image is available as supplementary material.


\section{Results and discussion}

\subsection{Field and forces}
The force field above the permanent magnet varies with distance to the
magnet surface in both strength and direction. For experiments with
magnetic nanoparticles inside the channel slide, we want the
lateral forces in the plane of the channel to be as small as possible,
yet the vertical force to be as high as possible and very uniform over
the area of interest.  Using the soft sphere model of
equation~(\ref{eq:chi}), we determined that the optimal height of
the channel slide for both conditions is \SI{10.5}{mm} above the
magnet.

\begin{figure}
    \centering
    \includegraphics[width=\figurewidth]{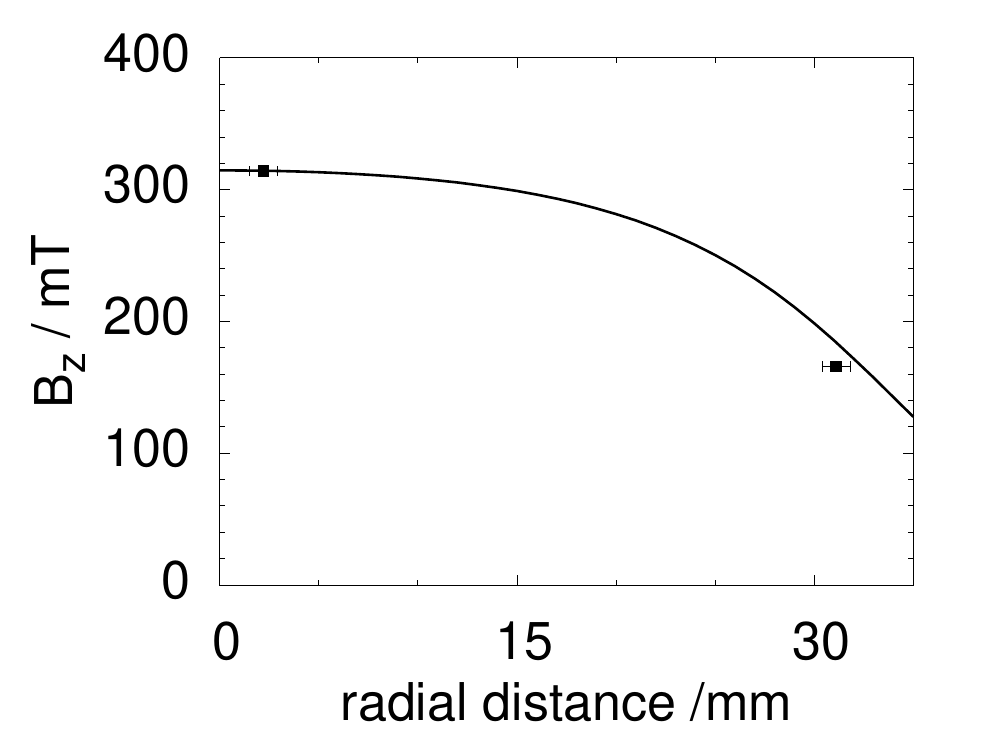}
        \includegraphics[width=\figurewidth]{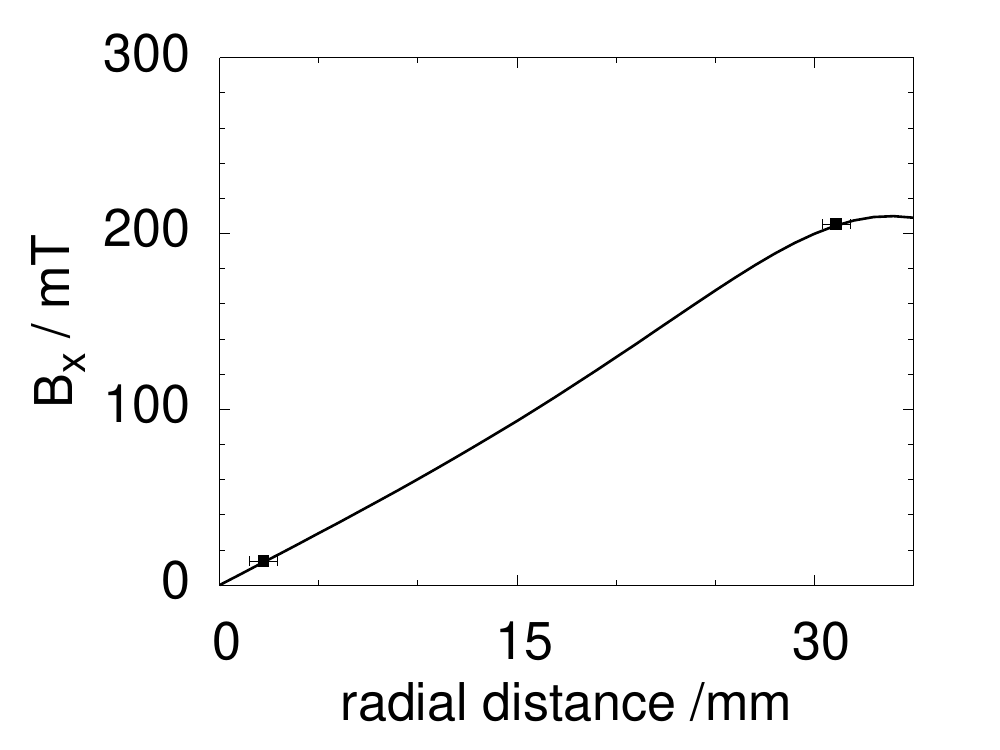}
        \caption{Calculated and measured vertical (top) and lateral
          (bottom) field components at the height level of the channel
          slide. Assuming a remanent magnetisation of \SI{1.27}{T},
          the field is predicted accurately.}
    \label{fig:Field}
\end{figure}

Figure \ref{fig:Field} shows the calculated magnetic field components
perpendicular to the channel ($B_\text{z}$) and along the channel
($B_\text{x}$) at this optimum height, compared with field measurements
in the center and at the edge of the magnet. A fit of the calculation
to the measurement results in a magnetisation of \SI{1.27(1)}{T}, which is
in agreement with the manufacturer's data. Over the length of the
channel (\SI{30}{mm}), the perpendicular field component varies by
\SI{5}{\percent}, and the field angle rotates by $\pm$\SI{17}{\degree}.

\begin{figure}
    \centering
    \includegraphics[width=\figurewidth]{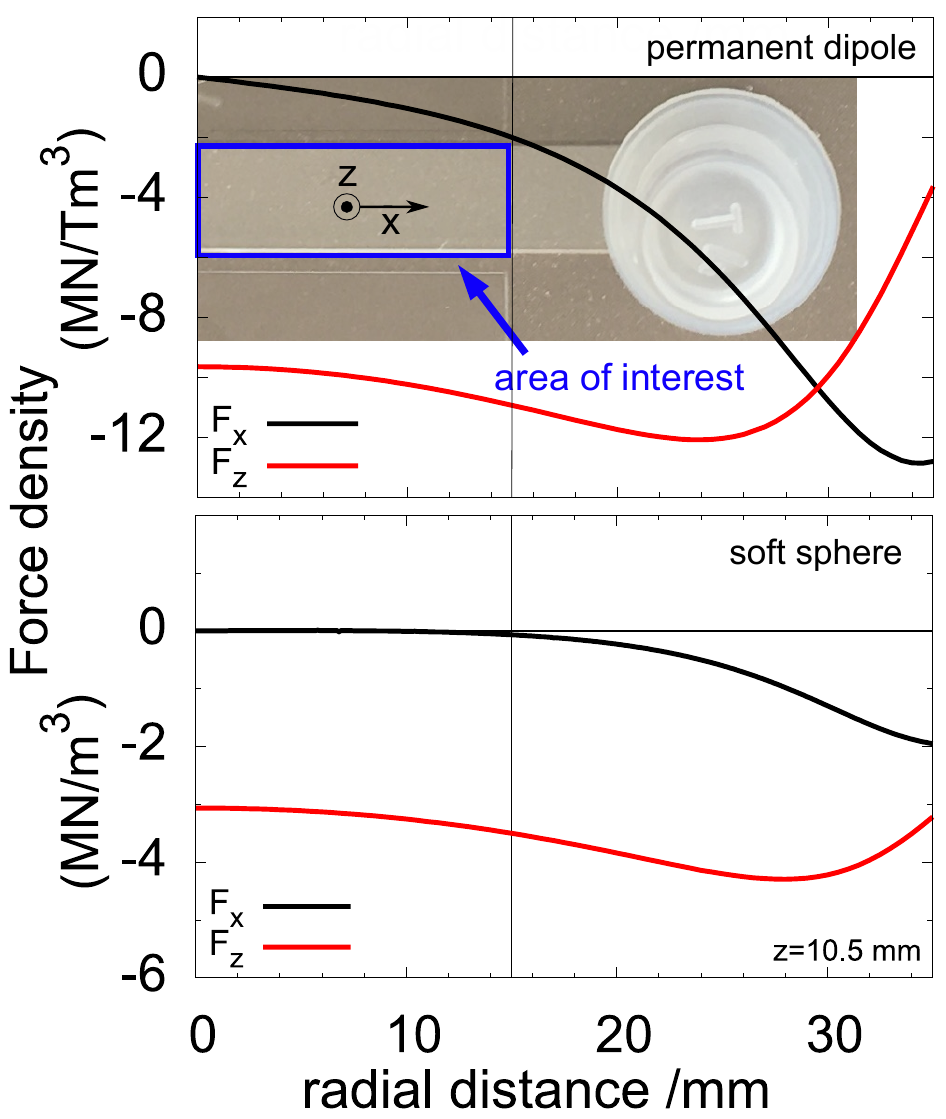}
    \caption{Calculated force densities in vertical (red) and lateral
      (black) direction as a function of the position over the length
      of the channel slide. The top curve indicates the case where the particle is a permanent magnet with remanent moment $I_\text{r}$
      (equation~\ref{eq:magnet}). To obtain the force, multiply by
      $V_\text{p}I_\text{r}$. The bottom curve indicates the case where the particle
      is a sphere of volume $V_\text{p}$ and susceptibility $\chi$
      (equation~\ref{eq:chi}). To obtain the force, multiply the value on the
      vertical axis by $3V_\text{p}\chi/(3+\chi)$.  }
    \label{fig:FzFxAreaInterest}
\end{figure}

Figure~\ref{fig:FzFxAreaInterest} shows the forces on particles for
both the permanent dipole magnet (top) and soft sphere (bottom)
models.  For the permanent dipole model, the forces are normalised to
the particle magnetic moment $I_\text{r} V_\text{p}$ [Tm$^3$]. A
typical particle has a magnetisation in the range from \SI{0.1}{}
to \SI{1}{T}, so force densities are on the order of
\SI{1}{} to \SI{10}{MN/m^3}. For comparison, the gravitation force density on a
typical iron-oxide particle is \SI{40}{kN/m^3} (gravitation
acceleration times mass density difference with water).

For the soft sphere model, the forces are normalised to
$3V_\text{p}\chi/(3+\chi)$, which ranges from \SI{0}{} to
\SI{3}{}$V_\text{p}$. A typical iron-oxide particle has a susceptibility
 greater than 1~\cite{Yun2014a}, meaning that force densities are
in the same range as for the permanent dipole approximation.

Both models show a variation of \SI{14}{\percent} in the vertical
component of the force over the length of the \SI{30}{mm} channel. The
soft sphere model has a much weaker lateral force. At the entry of
the channel, the force only tilts about \SI{1}{\degree} inward,
whereas for the permanent dipole model, it tilts at \SI{10}{\degree}.

\begin{figure}
    \centering
    \includegraphics[width=\figurewidth]{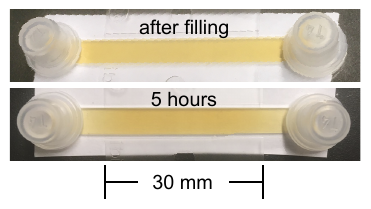}
    \caption{Channel filled with magnetic nanoparticles of \SI{5}{nm}
      diameter in water. Immediately after filling, the concentration
      is very uniform over the channel. Only after several hours can a
      slight increase in concentration be observed in the center. The concentration
      gradient is negligible over the region of interest.}
    \label{fig:Gradient5Hours}
\end{figure}

The large magnet produces a very uniform distribution of
magnetic particles. Figure~\ref{fig:Gradient5Hours} shows a channel
slide filled with a diluted suspension of \SI{5}{nm} iron-oxide
particles. Immediately after filling, no concentration gradient can be
observed. After several hours, a slight reduction in concentration is
observed close to the channel entries, yet 
the concentration appears optically uniform over the area of interest. Therefore, this magnet configuration applies forces that are at least
\SI{25}{} times higher than the gravitational force, perpendicular to
the surface and uniform over most of the channel slide.

\subsection{Cell viability}
\begin{figure}
    \centering
    \includegraphics[width=\widefigurewidth]{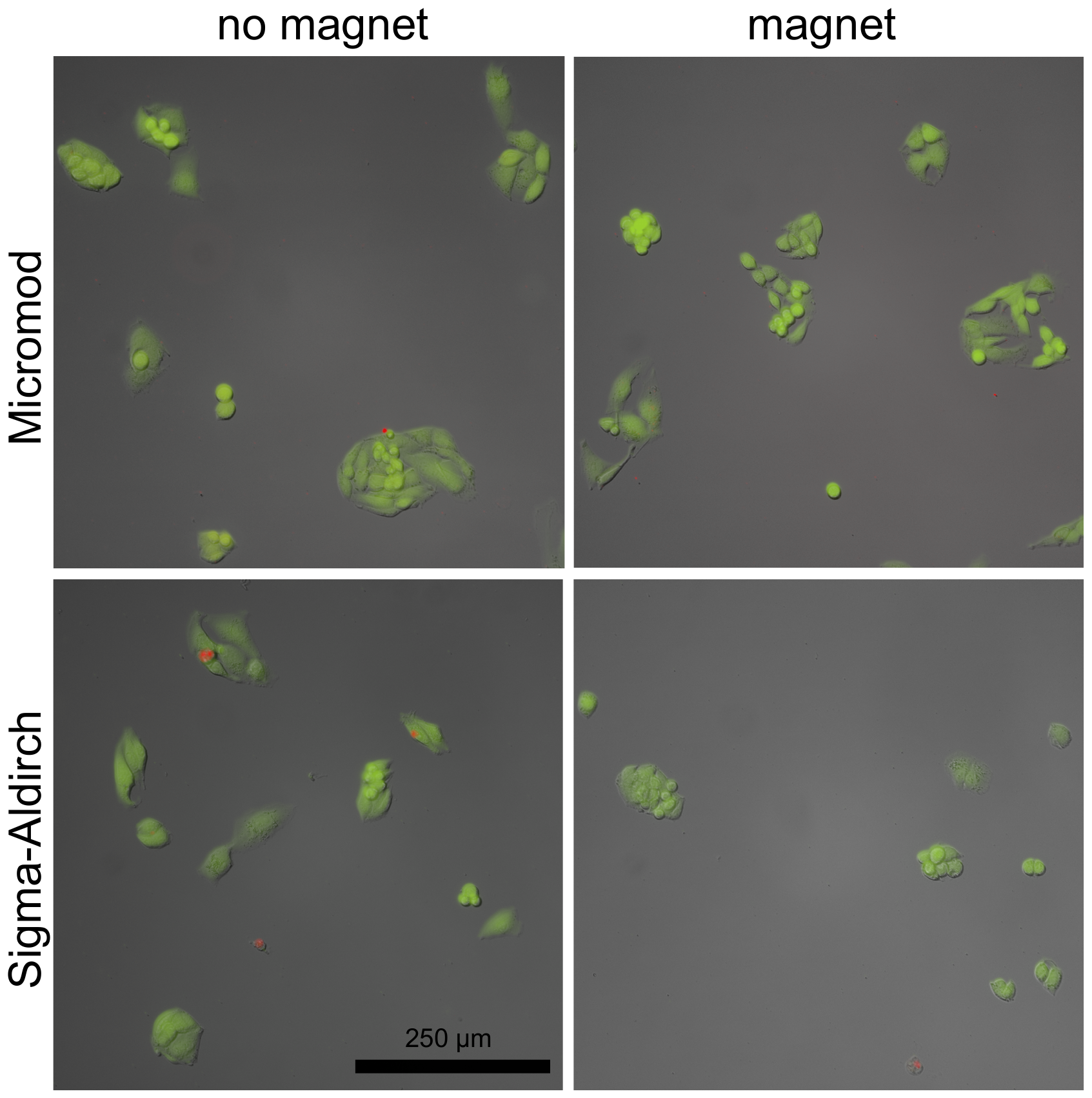}
    \caption{Composite images of a live/dead cell viability assay of
      \SI{5e4}{HepG2 cells/mL} exposed to \SI{5e8}{nanoparticles/mL}
      for \SI{24}{hours}. Dead cells are displayed in red, live cells
      in green. Top: Cells exposed to \SI{100}{nm} diameter
      cross-linked dextran iron-oxide composite particles
      (Micromod). Bottom: \SI{100}{nm} diameter iron oxide
      incorporated conjugated polymer particles (Sigma-Aldrich). The
      right-hand images are of cultures that were placed on the magnet. The
      cell survival for both types of particles and with or without
      field is higher than \SI{90}{\percent}, which is equal within
      measurement error to the control. Therefore, we conclude that
      these particles are non-toxic for a period of \SI{24}{hours}, as expected.}
    \label{fig:LiveDead}
\end{figure}

Two different fluorescent magnetic particles were used: cross-linked
dextran iron-oxide composite particles (from Micromod) and iron oxide
incorporated conjugated polymer particles (from Sigma-Aldrich). Both
particles have an average diameter of \SI{100}{nm}. These particles are expected to be non-toxic as they
were developed especially for cell studies. To confirm this, we performed a cell viability test on HepG2 cells
exposed for \SI{24}{hour} to approximately \SI{10000}{} particles per
cell. Experiments were performed both with and without application of
a magnetic field.  Figure~\ref{fig:LiveDead} shows a typical result
with a very small number of dead cells (red).  Analysis of thirty
images from three independent measurements shows that the survival
rate is higher than \SI{90}{\percent}, which is equal to the control
without particles within measurement uncertainty.

\subsection{Particle counting}
\begin{figure}
    \centering
    \includegraphics[width=\figurewidth]{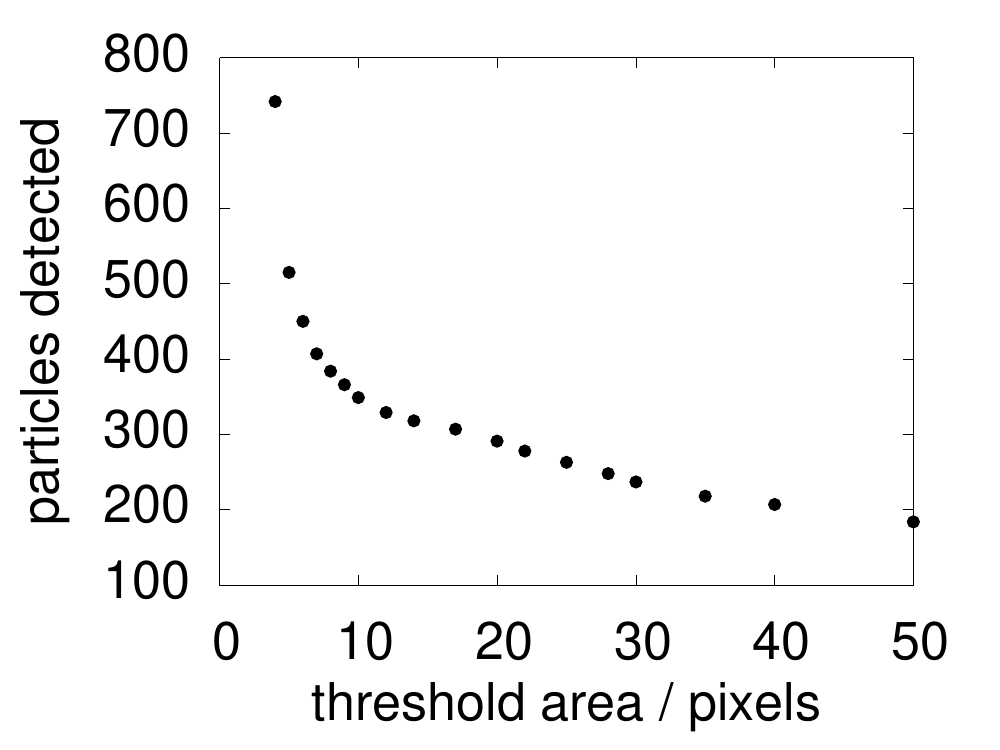}
    \includegraphics[width=\figurewidth]{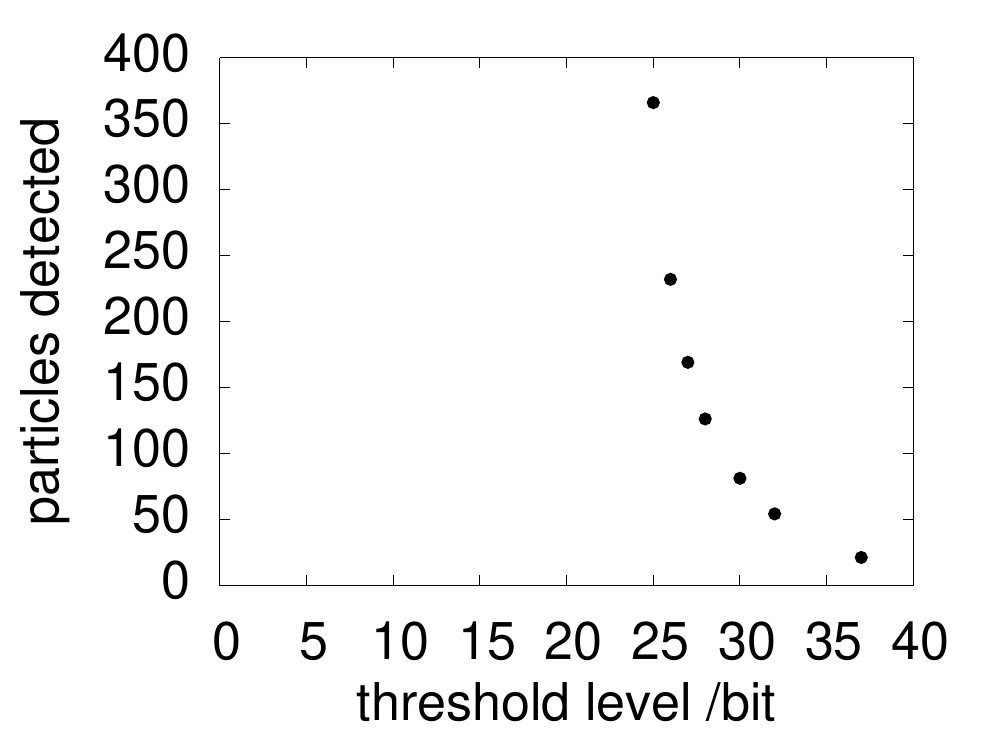}
    \caption{Number of detected red fluorescent particles as a
      function of the minimum size in pixels (top, using an intensity
      threshold of \num{25}) and detection threshold (bottom, using a
      size threshold of \SI{9}{pixels}). The number of detected
      particles is very sensitive to these settings, especially to the
      intensity threshold level.}
    \label{fig:Counts}
\end{figure}

Only the fluorescent particles can be imaged by means of an optical filter. To
eliminate observer bias, and for practical reasons, particle counting
was automated using ImageJ software (see Section~\ref{sec:exp}). The
image analysis procedure has two adjustable parameters. The first is
the threshold size of the observed particle in pixels. This value
should be sufficiently high to avoid false detection due to noise,
yet small enough not to miss particles. With a \num{20}$\times$ lens, the
pixel size is \SI{324}{nm}, which is lower than the wavelength of the
emitted light, so the minimum threshold size should be larger than
\SI{2x2}{pixel}. Figure~\ref{fig:Counts}, top image, shows that the
number of detected red fluorescent particles (Micromod) decreases with
increasing threshold size. Above \SI{3x3}{pixel}, the decrease is
linear. We assume that the rapid drop for small threshold sizes is caused
by noise. A reasonable value of \SI{3x3}{pixel} yields a value that is
about half the maximum value.

\begin{figure*}
    \centering
    \includegraphics[width=\textwidth]{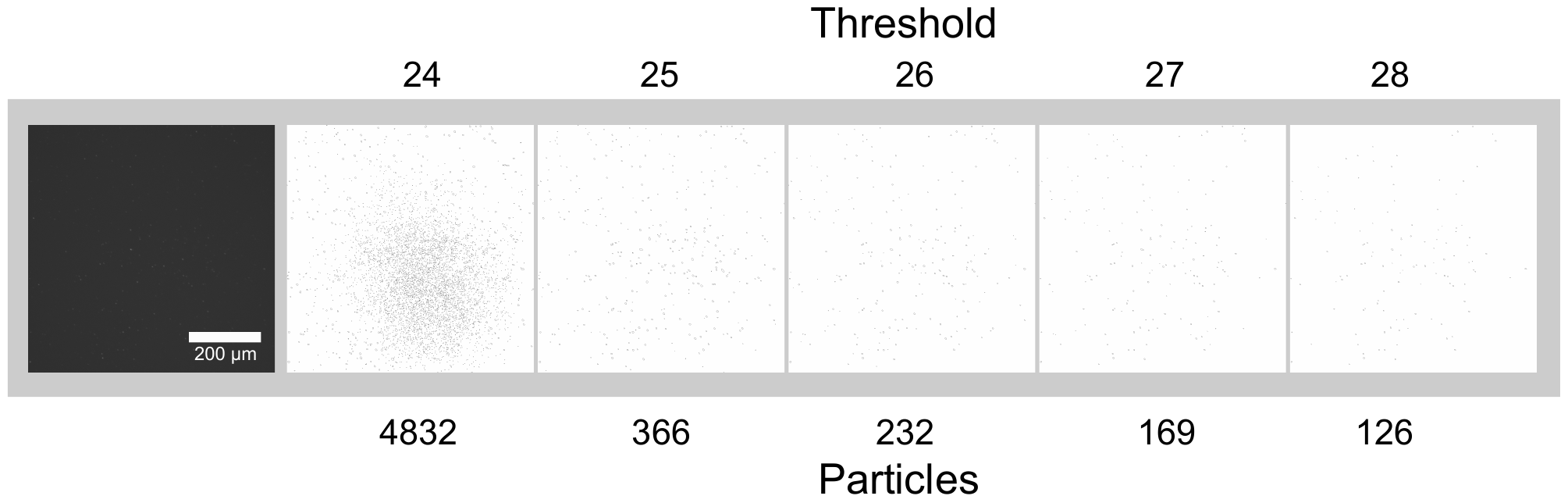}
    \caption{Image of Micromod particles taken with red filter
      (left) and detected particles using automated image analysis
      with a size threshold of \SI{9}{pixels} as a function of the
      intensity threshold (indicated above the images). The number of
      detected particles (indicated below the images) is very
      sensitive to the threshold chosen for optimal particle
      detection.}
    \label{fig:Threshold}
\end{figure*}

The second parameter is the intensity threshold, which discriminates
between the background and the particles. This value lies in the range
of \num{1}--\num{255}. It should not be so low that it avoids false detection by
noise, yet not so high that it misses faint particles. In contrast to the
size threshold, the amount of particles detected is extremely
sensitive to the intensity threshold, see the bottom graph of Figure~\ref{fig:Counts}. This is illustrated in the ImageJ results of the
particle detection algorithm shown in Figure~\ref{fig:Threshold}. At a
threshold of \num{24}, the non-uniform illumination disturbs the
image, whereas the output seems reasonable from \num{26} to \num{28}. However, the particle count  varies by a factor of \num{10}
over this small range.

The total number of particles expected from the concentration of
\SI{5e8}{particles/mL} is approximately \SI{1e5}{} in the field of
view. It is clear that, whatever the settings, the method grossly
underestimates the number of particles. This is very likely due to
particle agglomeration, but one should not rule out that particles 
lose their fluorescent properties. This does not render the method
entirely useless. By using an identical method to detect particles in
different images, a relative comparison can be made between the case with and
without the magnet. Therefore, we analysed images with a size
threshold of \SI{3x3}{pixel} and an intensity threshold as that value
at which the non-uniform illumination disappears (e.g.~\num{25} in
Figure~\ref{fig:Threshold}).

\subsection{Cell and particle count}
\begin{figure}
  \centering
  \includegraphics[width=\widefigurewidth]{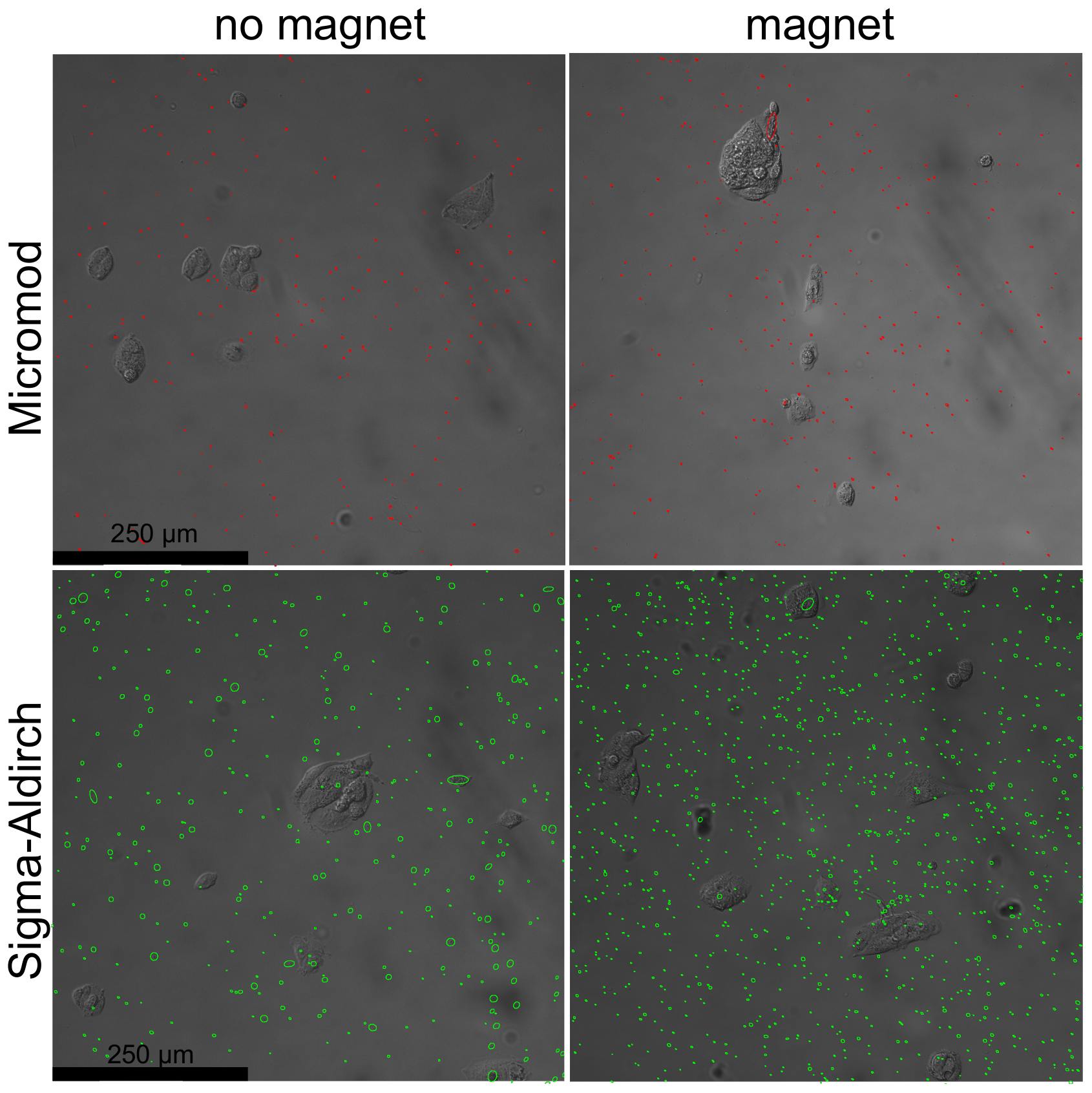}
  \caption{Composite image of HepG2 liver cells (greyscale) and detected
    Micromod red fluorescent particles (red, top row) and
    Sigma-Aldrich green fluorescent particles (green, bottom
    row). There is no significant difference between the sample
    that has been away from (left) or on top of
    (right) the magnet for \SI{24}{hours}.}
  \label{fig:BNFSigmaComposite}
\end{figure}

\begin{figure}
    \centering
    \includegraphics[width=\figurewidth]{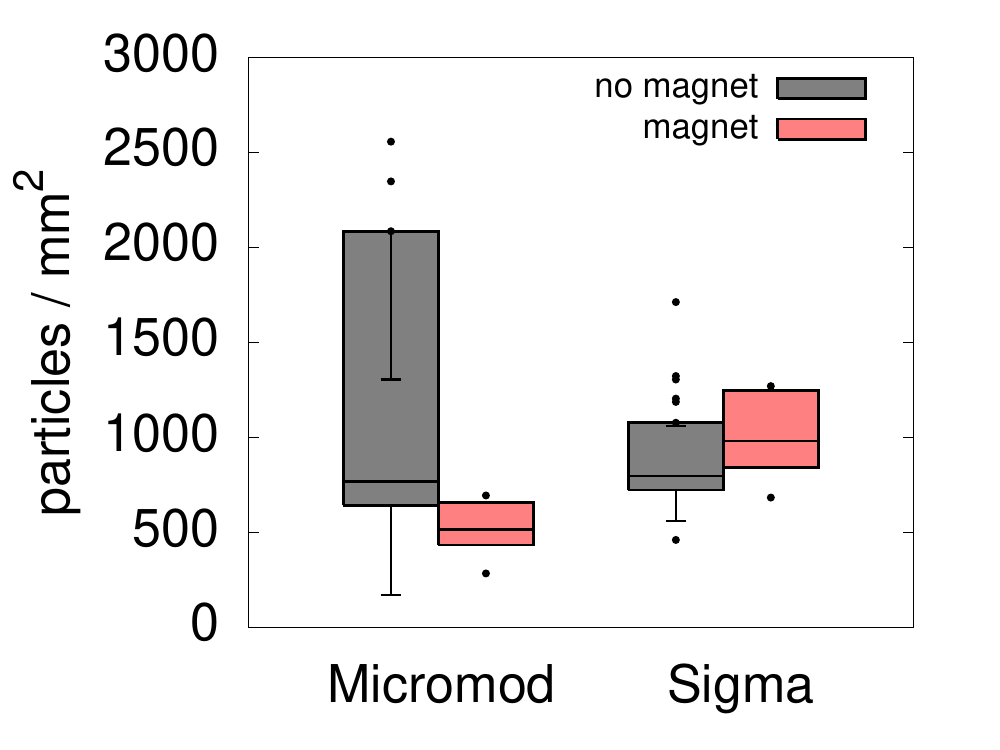}
    \caption{Particle density obtained from image analysis for samples
      with two different magnetic particles, away from (grey) and on
      top of (red) the magnet.  A box is drawn around the region
      between the first and third quartiles, with a horizontal line at
      the median value. Whiskers extending from the box encapsulate
      two-thirds of the data points. Data outside that range is
      indicated by points. There is no statistically significant
      difference in the four cases. For the Micromod particles,
      there is a hint of a reduction in particle density when the
      sample was on the magnet.}
    \label{fig:Particles}
\end{figure}

The image analysis method was applied to human liver cell cultures
with Micromod and Sigma-Aldrich particles, both with and without
magnetic field.  Figure~\ref{fig:BNFSigmaComposite} shows composite
images of typical results for these four cases. We superimposed the location of
detected particles in red (Micromod particles) and green
(Sigma-Aldrich particles) on top of the
greyscale bright field images. The cell and particle densities appear more
or less similar. We performed statistical analyses of \num{61} images
(for Micromod \num{17} and \num{7}, for Sigma-Aldrich \num{23} and
\num{14}, away from and on top of the magnet,
respectively). Figure~\ref{fig:Particles} shows there is no clear
difference in particle density among the four cases. There is a
hint that the particle density is lower in the experiment with the
MicroMod particles (left) on the magnet (red). Given the uncertainty
in the particle detection algorithm, however, we refrain from drawing a
definite conclusion.

\begin{figure}
    \centering
    \includegraphics[width=\figurewidth]{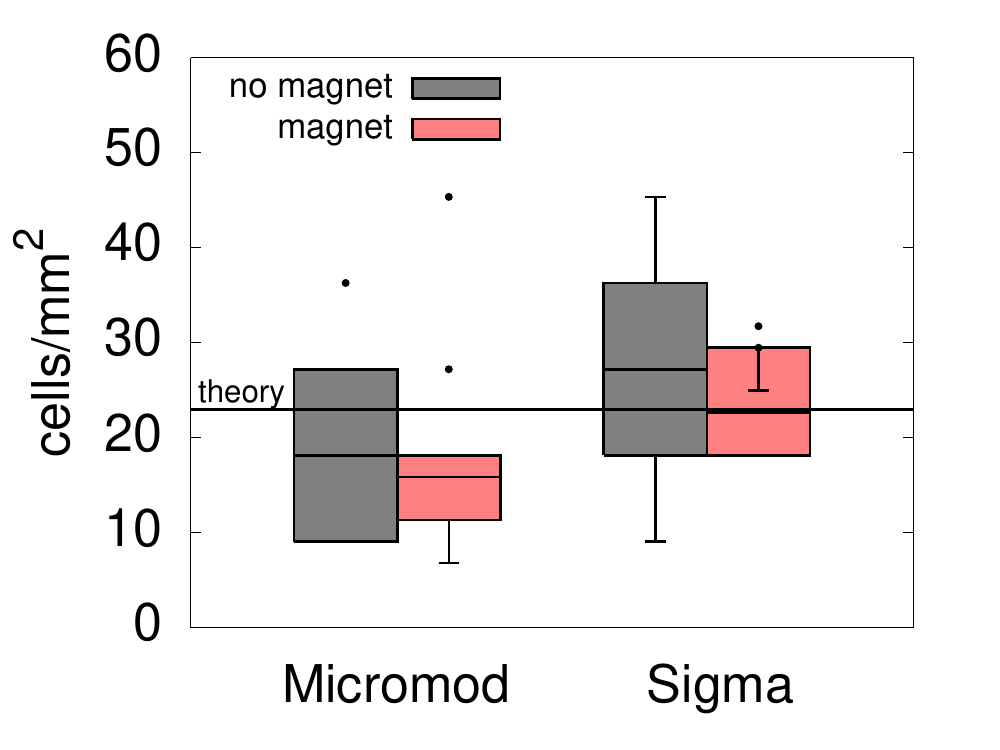}
    \caption{Observed cell density for samples with two different
      magnetic particles, away from (grey) and on top of (red) the
      magnet. The cell density is on the order of the
      expected value (black line labeled ``theory'') and there is no
      significant influence of the magnetic field.}
    \label{fig:Cells}
\end{figure}

In addition to particles, we also analysed the number of
cells. Figure~\ref{fig:Cells} shows the observed cell density for the
four different cases. As the particles do not affect cell
viability, we do not expect major differences. Indeed, the cell
density does not exhibit a statistically significant dependency on
particle type or field condition. The cell density is in agreement with
the cell concentration in the administered solution
(Section~\ref{sec:exp}), indicated by the black line labeled ``theory''.

\begin{figure}
    \centering
    \includegraphics[width=\figurewidth]{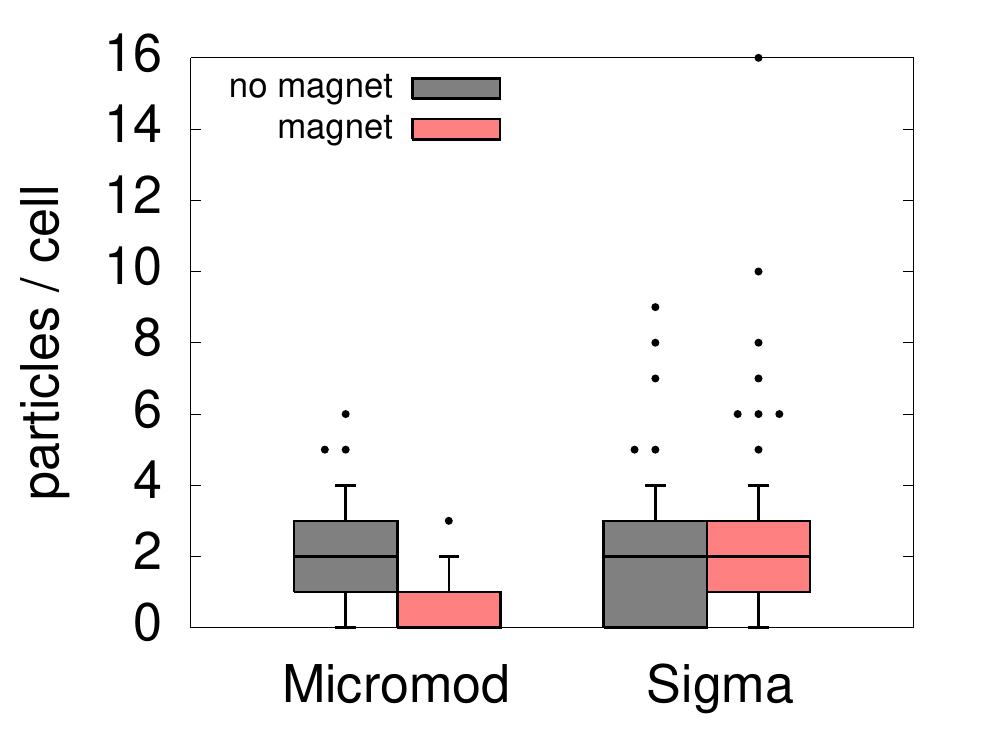}
    \caption{Number of particles that coincide with a cell for two
      different magnetic particles, away from (grey) and on top of
      (red) the magnet. The differences are not significant. As in the
      particle count  for the Micromod
      particles (Figure~\ref{fig:Particles}), there might be some reduction in particles per cell
      under application of a field.}
    \label{fig:ParticlesPerCell}
\end{figure}

We also counted the number of particles registered with cells in the
bright field image. Figure~\ref{fig:ParticlesPerCell} shows that, again,
there is no clear difference between the cells that were away
from and those on top of the magnet. However, it is surprising that we
observed less than ten particles on top of cells, which is far too
few. From the cell and particle densities shown in
Figures~\ref{fig:Particles} and \ref{fig:Cells}, one would expect
approximately \num{40} particles per cell. A possible explanation
could be that particles do not adhere well to the cell membrane. Nevertheless, one
should not rule out that particles may be incorporated into the
cell and lose their fluorescent properties.

From these observations we conclude that the type of particles used in
this experiment is not very suitable to determine whether there is an
enhanced uptake of magnetic particles under application of a strong
field gradient. Compared to the uncertainties caused by the image
analysis and spread between individual images, the differences between
the cases are small. This uncertainty prohibits us from drawing robust conclusions. Rather, one will have to apply more
complicated methods such as measuring the iron content by mass
spectrometry, transmission electron microscopy or transfection.

\section{Conclusions}
We constructed a magnetic system that exerts a strong force on
fluorescent magnetic nanoparticles inside a channel slide with a
human liver cell culture.

By using a cylindrical magnet with a diameter of \SI{70}{mm}, the
vector components of the field and force in the vertical direction
were dominant, with a strength in excess of \SI{300}{mT} and
\SI{6}{MN/m^3}, respectively. Over the length of the \SI{30}{mm}
channel, the force strength remains within \SI{14}{\percent} and the
force direction varies by less than \SI{1}{\degree}.

HepG2 human liver cells exposed for \SI{24}{hours} showed no
significant change in cell viability when exposed to cross-linked
dextran iron-oxide composite particles (Micromod) or iron oxide
incorporated conjugated polymer particles (Sigma-Aldrich), both of
\SI{100}{nm} diameter and with a concentration of
\SI{10000}{particles/cell}.

The number of fluorescent particles detected by optical microscopy
depends by at least one order of magnitude on the settings of the
automated image detection algorithm. The number of detected particles
is especially sensitive to the intensity level threshold.

Analysis of over \num{60} images did not show an increase in the
number of observed magnetic particles overlapping with cells when a
magnetic force is applied. On the contrary, the particle density seems
to be a factor of \num{40} lower on cells than in regions without
cells.

From these measurements, we conclude that using fluorescent magnetic
nanoparticles to demonstrate enhanced particle uptake in magnetic
fields is far from trivial, and may not be the optimal approach.
 


\clearpage
\bibliographystyle{bst/apsrev_modified_doi}
\bibliography{paperbase}

\end{document}